\documentclass[amsmath,amssymb,prl,showpacs,twocolumn,floatfix]{revtex4}
\usepackage{graphicx}
\usepackage{dcolumn}

\begin{document}
\title{Chiral Brownian Heat Pump} 
\author{M. van den Broek}
\author{C. Van den Broeck}
\affiliation{Hasselt University, B-3590 Diepenbeek, Belgium}
\begin{abstract}
We present the exact analysis of a chiral Brownian motor and heat pump. Optimization of the construction predicts,  for a nanoscale device, frequencies of the order of kHz and cooling rates of  the order of femtojoule per second.
\end{abstract}

\maketitle

Brownian motors have been studied intensively since the early 1990s \cite{brownianmotors}. This interest coincided with developments in bioengineering and nano\-techno\-logy, where understanding and designing a motor in the shape of a small biological or artificial device is an important issue. Most of the motors investigated in this context are powered by chemical energy. Brownian motors driven by a temperature gradient  \cite{thermalbrownianmotors} have a fundamental appeal, since their operation is directly related to basic questions such as Carnot efficiency, Maxwell demons and the foundations of statistical mechanics and thermodynamics \cite{basicquestions}. 
The additional significance of the thermal Brownian motor comes from the recent observation that it can operate as a refrigerator \cite{refrigerator}, see also \cite{pekola}. In fact, this property is, at least in the regime of linear response, a direct consequence of Onsager symmetry: if a temperature gradient generates motion, an applied force will generate a heat flux. This principle is well known in its application to electro-thermal devices, displaying  the Peltier, Seebeck and Thompson effects \cite{callen}. At variance however with these macroscopic devices, rectification of nonequilibrium thermal fluctuations provide the driving mechanism for Brownian refrigeration. The latter become more prominent, and so do the resulting motor and cooling functions, as the apparatus becomes smaller. 

Previous models for the Brownian refrigerator assume translational motion of the engine \cite{refrigerator}. This construction obviously poses difficulties in its technological implementation, while the resulting friction is expected to lower the efficiency. In this letter, we present a chiral rotational model, in which these problems do not occur, and which has the extra benefit that the choice of the axis of rotation provides an additional parameter that can be optimized. A related question, that will also be addressed, is the optimal chiral shape of the engine. The observed optimized rates of rotation and heat transfer are found to be significantly larger than in the translational counterpart, suggesting the technological implementation of such devices.

\begin{figure}
\includegraphics[width=\columnwidth]{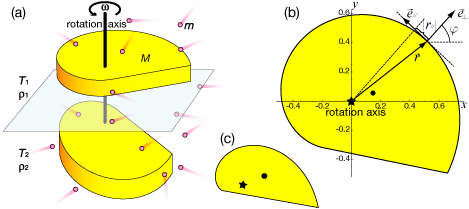}
\caption{\label{fig1}
(Color online)
(a) Schematic representation of the chiral motor (mass $M$) rotating as a single unit along the vertical $z$-axis. Each motor element resides in a separate compartment, filled with gas particles (mass $m$) at temperatures $T_{1}$, $T_{2}$ and  densities $\rho_{1}$, $\rho_{2}$, respectively.   
(b) The planar shape of one motor part is determined by the position vector $\vec{r}(x,y)$ of its perimeter (the origin being the axis of rotation).  The collision rule Eq.~(\ref{collisionlaw}) is given in terms of the polar angle $\varphi$, determining the direction orthogonal to the perimeter. The properties of the engine are expressed in terms of the tangential component $r_{\shortparallel} = \vec{r} \cdot \vec{e}_{\shortparallel}$ of $\vec{r}$ along the perimeter cf. Eqs.~(\ref{transitionprobability}, \ref{frictioncoefficient}-\ref{Q}).
Maximal rotation and refrigeration speed is attained for one motor unit being the enantiomorph of the other. The corresponding optimal shape (and axis implantation marked by a star) is the spiral form depicted here for maximum frequency (b) and maximum net cooling power (c).
}
\end{figure}
Since the properties of the Brownian heat pump follow by Onsager symmetry from those of the Brownian motor, we first focus on the latter. The basic construction is represented in Fig.\:\ref{fig1}(a). The engine
consists of two parts linked by a rigid axis (which we take to be the $z$-axis), around which the whole construction is free to rotate as a single entity.  This rotational motion is induced, following Newton's laws, by the collisions with surrounding gas particles. The question of interest is under which conditions sustained rotational motion will arise. Following the Curie principle, the breaking of symmetry plays a crucial role. There are two obvious symmetries involved:
the statistical symmetry of the microscopic dynamics
and the chiral symmetry of the device. If the motor units are achiral, in the sense that clockwise and counterclockwise rotation cannot be distinguished, no sustained motion will appear.  When both motor units reside in a single compartment at
equilibrium, sustained rotational motion will not appear, even for chiral motor units, because it would violate the second law of thermodynamics, or referring to the basic underlying symmetry, because detailed balance should hold \cite{onsager}.
Both symmetries will be broken if we consider
chiral units residing in  separate 
compartments at unequal temperatures $T_i$. The index $i$ runs over the different reservoirs, cf. Fig.\:\ref{fig1}(a) for a schematic representation in the case of two reservoirs $i=1,2$. As we proceed to show, the resulting average rotational frequency can be calculated exactly from microscopic dynamics, at least in a limiting case. 
For simplicity, we will restrict the theoretical analysis in this letter to the case of a two-dimensional device. The corresponding results for three-dimensional cylindrical objects, as depicted in  Fig.\:\ref{fig1}(a), are obtained by appropriate rescaling with the hight $h$ of the units
(for more details, see \cite{martijn}). 

Turning to the exact microscopic analysis, we consider convex (two-dimensional) units,  residing in reservoirs that are infinitely large and are filled with dilute gases at equilibrium. The reservoirs play the role of ideal thermostats with which the engine is exchanging energy.  In the limit of high dilution and a heavy engine (mass $M$, moment of inertia $I$), the collisions of the engine parts with the gas particles (mass $m$) become uncorrelated events and  the following exact Boltzmann master equation  for the probability distribution $P_t(\omega)$  of its angular velocity holds:
\begin{equation}\label{masterequation}
\frac{\partial P_t(\omega)}{\partial t}
= \int d\omega' \left[W_{\omega|\omega'} P_t(\omega') - W_{\omega'|\omega}  P_t(\omega) \right].
\end{equation}
Here $W_{\omega|\omega'}$ is the transition probability per unit time for the motor to
change its angular velocity from $\omega'$ to $\omega$ by one collision. We will assume that the collisions are  perfectly elastic and that the interaction force is short-ranged and central.  Conservation of the total energy and of the total angular momentum in the $z$-direction of the colliding pair, and of the tangential component of the momentum of the gas particle, leads to the following collision law: 
\begin{equation}\label{collisionlaw}
\omega
= \omega' +  \frac{2 (\omega' y + v'_{x}) \cos \varphi -  2 (\omega' x - v'_{y}) \sin \varphi}
{ x \sin \varphi - y \cos \varphi + \frac{I}{m} \left(x \sin \varphi - y \cos \varphi \right)^{-1}},
\end{equation}
specifying $\omega$ in terms of its pre-collisional value $\omega'$ and the pre-collisional speeds  $\vec{v'} = (v'_{x}, v'_{y})$ of the gas particle.
$\varphi $ is the polar angle of the surface at impact, see Fig.\:\ref{fig1}(b).

 The transition probability $W_{\omega|\omega'}$ can now be calculated following standard arguments from the kinetic theory of gases. Taking into account that  the velocity distributions of the particles
$\phi_{i}(\vec{v})$ are Maxwellian at the density $\rho_i$ and temperature $T_i$ of their bath, one finds the following explicit expression:
\begin{multline}\label{transitionprobability}
W_{\omega|\omega'}
= \tfrac{1}{2 \pi} \sum_{i}
\oint dl_{i} 
\frac{\rho_{i}}{v_{T_{i}}}
( r_{\shortparallel} + \frac{I}{m r_{\shortparallel}})^{2}
|\Delta \omega|
H[r_{\shortparallel} \Delta \omega]
\\ 
\times 
 \exp \left[- 
 \left((r_{\shortparallel} + \frac{I}{m r_{\shortparallel}}) \Delta \omega + 2 r_{\shortparallel} \omega' \right)^{2} / \left(\pi v_{T_{i}}^{2}\right)
\right ].
\end{multline}
The contour integral $\oint dl_{i}$ is over the perimeter of the engine part in each reservoir, which can be of any convex shape. We have introduced the tangential component of the position vector $\vec{r}$ along the perimeter, $r_{\shortparallel} = \vec{r} \cdot \vec{e}_{\shortparallel} = - x \sin \varphi + y \cos \varphi$ cf. Fig.\:\ref{fig1}(b),
the thermal speed of the gas particles $v_{T_{i}} = \sqrt{8 k T_{i} / \pi m}$, and the angular velocity increment $\Delta \omega = \omega - \omega'$.

In equilibrium, $T_i = T$, one easily verifies that the Boltzmann distribution 
$P^{eq}(\omega)= \sqrt{I / (2 \pi k T)} \exp[- I \omega^{2} / (2 k T)]$
is the unique steady state solution of Eq.~(\ref{masterequation}). As required by statistical mechanics, this solution satisfies detailed balance: $W_{\omega|\omega'} P^{eq}(\omega')=W_{-\omega'|-\omega} P^{eq}(-\omega)$.

A general solution away from equilibrium is not available, hence we resort to a  perturbational technique. For small values of the parameter $\varepsilon =\sqrt{m /M}=r_{0} \sqrt{m/I}$, where $r_{0} = \sqrt{I /M}$ is the radius of gyration of the device, the change in 
angular velocity upon collision with a gas particle is small and we can apply a Kramers-Moyal type of expansion. Note that this expansion is not uniform when performed in terms of the probability distribution, but has been found to converge very well when performed at the level of the moments. Here we only present the final results of this procedure, for more details see \cite{martijn}.

Up to first order in $\varepsilon$ the average  angular speed of the motor obeys the following equation: 
\begin{multline} \label{expansionomega}
\frac{\partial \langle \omega  \rangle}{\partial t} 
= 
\frac{m}{M}
\sum_{i} \rho_{i} 
 \biggl[
 - v_{T_{i}} \langle \omega \rangle  \oint dl_{i} \left(\frac{r_{\shortparallel}}{r_0} \right)^{2}
\\
+ \varepsilon \left(
\sqrt{\frac{I}{m}} \langle \omega ^{2} \rangle
-\frac{k T_{i}}{\sqrt{mI}}\right) \oint dl_{i} \left(\frac{r_{\shortparallel}}{r_0} \right)^{3}
+ O(\varepsilon^{2})
 \biggr]. 
\end{multline}
To lowest order in $\varepsilon$ we recognize a linear drag law, 
$I \partial_{t} \langle \omega \rangle = - \gamma \langle \omega \rangle$,
featuring a frictional torque  which is proportional to the average rotational speed. The proportionality factor $\gamma = \sum_{i} \gamma_{i}$ is equal to the sum of the friction coefficients $\gamma_{i}$ contributed by each of the engine parts. From Eq.~(\ref{expansionomega}), one finds the following explicit expression for these friction coefficients:
\begin{equation}\label{frictioncoefficient}
\gamma_{i} 
= m \rho_{i} v_{T_{i}} \oint dl_{i} r_{\shortparallel}^{2}.
\end{equation}
Note that at this order of the perturbation, no systematic steady state motion appears $\langle \omega
\rangle^{st}=0$. This indicates that the ``rectification of the fluctuations''  leading to systematic motion appears at the level of nonlinear and non-Gaussian effects. 

\begin{table*}
\caption{\label{table1}
Properties of the 3-d device with a shape optimized for maximum rotational frequency and cooling power, respectively (see Fig.\:\ref{fig1}). Each unit of the device is cylindrical with parallel surfaces of area $\pi R^{2}$, 
$R = 3 \thinspace$nm, height $h = 3\thinspace$nm. We assume following  values of the parameters: density  1350$\thinspace \text{kg}/\text{m}^{3}$ (typical for proteins); total mass  $M = 2.29 \times 10^{-22} \thinspace$kg; mass ratio  $m/M = 1.3 \times 10^{-4}$ ($m$ being the mass of a water molecule);  $T = 300 \thinspace$K; temperature gradient for the motor: $\Delta T = 0.1 \thinspace$K.
}
\begin{ruledtabular}
\begin{tabular}{l c c c c  c c}
& $\langle \omega \rangle$ (Hz) & 
$\gamma$ ($ 10^{-28} \thinspace \text{Nms}$)&
$I$ ($ 10^{-39} \thinspace \text{kg} \thinspace\text{m}^{2}$)&
$\dot{Q}_{1 \rightarrow 2} / \Gamma$ ($ 10^{6} \thinspace$J/(Nms)) &
$\Gamma_{\text{lim}}$ ($ 10^{-21} \thinspace \text{Nm}$) &
$\dot{Q}_{\text{net}}^{\text{max}}$ ($ 10^{-15} \thinspace \text{J}/\text{s}$) 
\\
\hline
Motor & 2180 & 0.90 & 1.26 & 6.53 & 1.17 & 1.92 \\
Heat pump & 1470 & 4.55 & 2.22 & 4.41 & 4.01 & 4.42 
\end{tabular}
\end{ruledtabular}
\end{table*}

At the next order in $\varepsilon$, the equation for the first moment is  coupled to the second moment, whose evaluation is thus needed to close the equation. Restricting ourselves to the steady state, one finds, not surprisingly, that  (to lowest order) the average kinetic energy of the motor is given by the usual expression for equipartition,
$\frac{1}{2} I \langle \omega^{2} \rangle = \frac{1}{2} k T_{\text{eff}}$, but at an effective temperature $T_{\text{eff}}$. The latter is found to be equal to the weighted geometric mean of the temperatures in the reservoirs:
$T_{\text{eff}} = (\sum_i{\gamma_{i} T_i}) / (\sum_i{\gamma_{i}})$.

Combined with Eq.~(\ref{expansionomega}) we conclude that (up to first order $\varepsilon$), the engine will develop an average steady state angular velocity given by
 \begin{equation}\label{omega}
\langle \omega \rangle
= 
\frac{\sum_{i} \rho_{i} k (T_{\text{eff}} - T_{i}) \oint dl_{i} r_{\shortparallel}^{3}}
{I \sum_{i}\rho_{i} v_{T_{i}}
\oint dl_{i} r_{\shortparallel}^{2}}.
\end{equation}

\begin{figure}[b]
\includegraphics[width=\columnwidth]{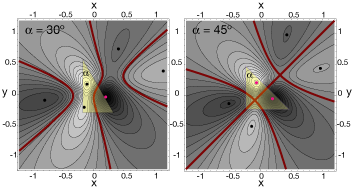}
\caption{\label{fig2}
(Color online)
Equal-amplitude lines of the rotational frequency $\langle \omega \rangle$
 as a function of the location of the rotational axis, for two triangular shapes (apex angle $30^{\circ}$ and $45^{\circ}$). The $(x,y)$ coordinates represent the location of the axis with respect to the center of mass of the unit. Thick lines correspond to $\langle \omega \rangle = 0$, and dots with maximum frequency.
}
\end{figure}
We proceed to discuss this first central result of our paper.  The angular velocity is obviously zero at equilibrium, $T_i=T_{\text{eff}}$, and  also when
$\oint dl_{i} r_{\shortparallel}^{3} = 0$, in agreement with the fact that the object then loses its chirality.
As far as maximizing rotational frequency is concerned,
a numerical procedure was employed to identify the optimum configuration (shape plus axis implantation) by deforming the contours in both compartments.
This resulted in the spiral shape depicted in Fig.\:\ref{fig1}(b), with one engine part the enantiomorph of the other one.
The same shape remains optimal, but appearing as the basis of a cylindrical object as depicted in  Fig.\:\ref{fig1}(a), when turning to the  case of dimension 3.
The dependence of the rotational speed on the shape and axis implantation is very intricate. In fact even the direction of the net rotation is not at all obvious. In  Fig.\:\ref{fig2} we reproduce,  for a specific triangular motor element (and its enantiomorph), the lines of equal amplitude  for the rotational frequency as a function  of the implantation of the rotation axis. The rotation is clockwise/counterclockwise in the dark/light shaded areas respectively.
In view of the technological interest of this result, we include the corresponding properties of such engine calculated under physically realistic conditions, in Table\:\ref{table1}. Rotational frequencies in the kHz regime are obtained for a temperature gradient of .1 K. Such a sustained average rotation will dominate over the thermal Brownian motion, with typical frequency $\sqrt{k T/I} = 1.81\times10^{9} \thinspace$Hz, on a time scale of seconds or more.

We next turn to the analysis of the heat pump function.  In the following, we will focus only on the linear response property, which can directly be obtained by invoking Onsager symmetry. To do so, one needs to write the result Eq.~(\ref{omega}) in the framework of linear irreversible thermodynamics~\cite{callen}. One identifies the flux $J_1 = \langle \omega \rangle$ and  the thermodynamic force $X_2 = 1/T_2-1/T_1$. For a small temperature
difference $\Delta T$, $T_1=T+\Delta T/2$, $T_2=T-\Delta T/2$,  a linear
relation  between flux $J_1$ and force $X_2 = \Delta T/T^2$ is observed, namely:  $J_1 = L_{12} X_2$.
The value of the coefficient $L_{12}$ is found from Eq.~(\ref{omega}) (for simplicity considering again enantiomorphs):
\begin{equation}\label{L12}
L_{12} 
= 
\frac{2 k T^{2}}{I v_{T}}
\frac{\rho_{1} \rho_{2}}{(\rho_{1} + \rho_{2})^{2}} 
\frac{\oint dl\,  r_{\shortparallel}^{3}}{\oint  dl\,  r_{\shortparallel}^{2}}.
\end{equation}
Following Onsager symmetry~\cite{onsager}, there is a mirror relation $J_2 = L_{21}  X_1 $
with an identical proportionality coefficient  $L_{21} = L_{12}$, while  $J_2$ is the flux associated to the temperature gradient $X_2$,
i.e.,  it is a heat flux $\dot{Q}_{1 \rightarrow 2}$  (from reservoir 1
to reservoir 2), and $X_1$ is the thermodynamic force associated with the rotation, namely a mechanical 
torque  divided by the temperature of the system, $X_1=\Gamma/T$. 
The relation $J_2 = L_{21} X_1$,  with  Eq.~(\ref{L12}),
implies that the heat flux $\dot{Q}_{1 \rightarrow 2}$  is given by
\begin{equation}\label{Q}
\dot{Q}_{1 \rightarrow 2}
= 
\frac{2 k T}{I v_{T}}
\frac{\rho_{1} \rho_{2}}{(\rho_{1} + \rho_{2})^{2}} 
\frac{\oint dl\,  r_{\shortparallel}^{3}}{\oint  dl\,  r_{\shortparallel}^{2}}
\, \Gamma.
\end{equation}
This is the second basic result of this letter. Note  that the direction of heat transfer depends on the direction of the torque, in such a
way that it activates an opposing Brownian motor in agreement with Le Chatelier's principle \cite{callen}.  For example, considering $\oint dl\,  r_{\shortparallel}^{3}>0$, the motor rotates clockwise $\langle \omega \rangle<0$ for $\Delta T<0$ ($ T_1<T_2$). The application of a positive torque $\Gamma >0$, inducing counterclockwise rotation, produces an energy flux $\dot{Q}_{1 \rightarrow 2}>0$, tending to activate the clockwise Brownian motor.

\begin{figure}[t]
\includegraphics[width=\columnwidth]{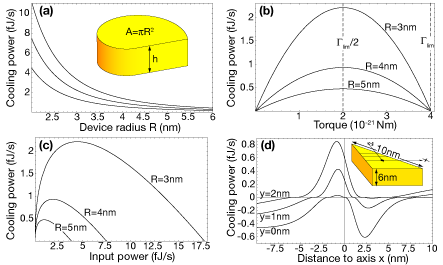}
\caption{\label{fig3}
(Color online)
The cooling power as a function of
(a) the radius of the device for heights $h = 4, 6 , 10 \thinspace$nm (half limiting torque, higher curve corresponds with lower $h$),
(b) the applied torque $\Gamma$ for height $h = 6 \thinspace$nm and given radius $R$, 
(c) the input power $P_{\text{in}}$ for the same dimensions, and
(d) the location of the rotation axis with respect to the center of mass.
The shape of the device is optimal for (a,b,c), while for (d) it corresponds with the $45^{\circ}$ configuration of Fig.\:\ref{fig2}.
Other properties as for Table~\ref{table1}.
}
\end{figure}

The Onsager coefficients $L_{21}$ and $L_{12}$ are the off-diagonal elements of the 2$\times$2 linear response matrix:
\begin{align}
\label{onsagermatrix}
J_{1} &= L_{11} X_{1} +  L_{12} X_{2}, &
L_{11} &= T / \gamma, 
\nonumber \\
J_{2} &= L_{21} X_{1} +  L_{22} X_{2}, &
L_{22} &= \gamma_{1} \gamma_{2} k T^{2} / (\gamma I).
\end{align}
The diagonal elements $L_{11}$, the rotational mobility, and $L_{22}$, the thermal conductivity, can again be calculated from the above perturbational method \cite{martijn}, or from general arguments based on Langevin theory~\cite{basicquestions}.  
As we proceed to show, these terms, associated with Joule heating and heat conduction, specify the domain in which the heat pump can operate as a cooling device.

An external torque can induce a cooling flux, by pumping heat out of one reservoir (into the other).
This effect is offset by
a dissipative contribution in both reservoirs,  resulting from frictional heating.
The linear response term $L_{11} X_{1}$ expresses that work is performed on the pump upon application of an external torque $\Gamma$, which leads to a power input $ \Gamma^{2} \gamma_{i} / \gamma^{2}$ in each reservoir $i$.
The $\Gamma^{2}$ dependency ensures that the cooling flux, which is proportional to $\Gamma$, will dominate for torques below a certain $\Gamma_{\text{lim}}$, as depicted in Fig.\:\ref{fig3}(b) for a concrete realization of the device.
The formal condition $|L_{21} X_1|> \Gamma^{2} \gamma_{1} / \gamma^{2}$ enables us to quantify the limiting torque as $\Gamma_{\text{lim}}=\gamma^{2} |L_{12}|/(\gamma_1 T)$,
which is remarkably scale-independent.
Maximum net cooling occurs for half the limiting torque $\Gamma_{\text{lim}} / 2$. Under this condition, a device of a few nanometer thickness is capable of a net rate of femtojoules per second, cf. Fig.\:\ref{fig3}(a) and Table\:\ref{table1}.
We also note that under this condition, half  of the input power $P_{\text{in}}$ is used for cooling, yielding a coefficient of performance $\eta =  \dot{Q}_{\text{net}} / P_{\text{in}} = 0.5$. For lower values of the torque, $\Gamma < \Gamma_{\text{lim}} / 2$, a higher  performance $\eta$ is feasible even though, as explained above, the net cooling flux is no longer maximal [see also Fig.\:\ref{fig3}(c)]. 

We finally turn to the issue of thermal conductivity. Suppose that a temperature gradient develops under the application of an external torque, cooling one reservoir and heating the other. The heat pump, being in contact with reservoirs of unequal temperature, will then conduct heat against the cooling flow.  Eq.~(\ref{onsagermatrix}) tells us that this heat flow, $J_{2} = L_{22} X_{2}$, has the form of a Fourier law, $\kappa \Delta T$, with conductivity
$\kappa = L_{22} / T^{2}$.
An upper limit for the relative gradient $\Delta T/T$ emerges when the conductive flow $L_{22}X_2$ equals the cooling power $|L_{21} X_1|$.
In explicit terms and at maximum performance, $\Gamma=\Gamma_{\text{lim}}/2$, the temperature gradient is bounded by
\begin{equation}\label{tempgradient}
\frac{\Delta T}{T}
 =
 \frac{\pi}{8}\frac{m}{I} \frac{\rho_{2}}{\rho_{1} + \rho_{2}}
\frac{\left(\oint dl\,  r_{\shortparallel}^{3}\right)^{2}}
{\left(\oint dl\,  r_{\shortparallel}^{2}\right)^{2}}.
\end{equation}
The fact that this term is proportional with $m/M$ indicates that the device may be better suited to transfer heat than to create a direct temperature difference.

In conclusion, a chiral molecular device, operating as a heat engine or heat pump, appears to be technologically feasible. It remains to be seen whether the simplifications assumed in this exact theoretical analysis (ideal gas reservoirs, frictionless rotation axis and lowest order approximation in $m/M$) lead to a realistic estimation of the performance, and whether alternative constructions (non-rigid coupling between units, vibrational instead of rotational units) offer an even better perspective.

\end{document}